\documentclass[journal,10pt,draftclsnofoot,onecolumn]{IEEEtran}
\usepackage{amsmath,amssymb,graphicx}
\usepackage{epsfig}
\usepackage{psfrag}

\newtheorem{remark}{Remark}
\newtheorem{theorem}{Theorem}

\newtheorem{corollary}{Corollary}

\newcommand{\nb}{\bar{n}}
\newcommand{\pam}{\frac{\sqrt{\l nP}}{2L}}
\newcommand{\bl}{{\biggl(}}
\newcommand{\blb}{{\bigl\{}}
\newcommand{\br}{{\biggr)}}
\newcommand{\brb}{{\bigr\}}}
\newcommand{\suchthat}{\colon}

\usepackage{xspace}
\usepackage{bbm}
\usepackage{mathrsfs}

\newcommand{\Ps}{{\sf P}}
\newcommand{\Es}{{\sf E}}

\newcommand{\Ec}{\mathcal{E}}

\newcommand{\Wh}{{\hat{W}}}
\newcommand{\Xh}{{\hat{X}}}

\newcommand{\wh}{{\hat{w}}}

\newcommand{\Wt}{{\tilde{W}}}

\newcommand{\Yt}{{\tilde{Y}}}
\newcommand{\Zt}{{\tilde{Z}}}

\newcommand{\wt}{{\tilde{w}}}

\newcommand{\yt}{{\tilde{y}}}

\def\a{\alpha}

\def\d{\delta}

\def\eps{\epsilon}
\def\l{\lambda}

\let\P\relax
\DeclareMathOperator\P{\sf P}

\def\error{\mathrm{e}}

\newcommand{\N}{\mathrm{N}}

\def\textiid{i.i.d.\@\xspace}
\newcommand\iid{\ifmmode\text{ i.i.d. } \else \textiid \fi}

\ifCLASSINFOpdf
\else
\fi

\begin{document}
%
\title{Gaussian Channel with Noisy Feedback and Peak Energy Constraint}

\author{Yu~Xiang and Young-Han~Kim%
\thanks{The authors are with the Department
of Electrical and Computer Engineering, University of California, San Diego, La Jolla, CA 92093 USA
 e-mail: (yxiang@ucsd.edu; yhk@ucsd.edu).}
}



\maketitle

\begin{abstract}
Optimal coding over the additive white Gaussian noise channel under
the {\em peak} energy constraint is studied when there is noisy feedback
over an orthogonal additive white Gaussian noise channel.
As shown by Pinsker, under the peak energy constraint, the best error exponent for communicating
an $M$-ary message, $M \ge 3$, with noise-free feedback is strictly larger than the one without feedback.
This paper extends Pinsker's result and shows that if the noise power in the
feedback link is sufficiently small, the best error exponent
for conmmunicating an $M$-ary message can be strictly larger than
the one without feedback. The proof involves two feedback coding schemes.
One is motivated by a two-stage noisy feedback coding scheme of Burnashev and Yamamoto for binary symmetric channels, while the other is a linear noisy feedback coding scheme that extends Pinsker's noise-free feedback coding scheme. When the feedback noise power $\a$ is sufficiently small, the linear coding scheme outperforms the two-stage (nonlinear) coding scheme, and is asymptotically optimal as $\a$ tends to zero. By contrast, when $\a$ is relatively larger, the two-stage coding scheme performs better.
\end{abstract}


\section{Introduction and Main Results}

\IEEEPARstart We consider a communication problem for an additive white Gaussian
noise (AWGN) \emph{forward} channel with feedback over an orthogonal
additive white Gaussian noise \emph{backward} channel as depicted in
Fig.~1.
\begin{figure}[htbp]
\begin{center}
\footnotesize\rmfamily
\psfrag{M}[b1]{\hspace{1em}Encoder}
\psfrag{N}[b1]{Decoder}
\psfrag{B}[b1]{\hspace{1em}$X_i$}
\psfrag{E}[b1]{\hspace{1em}$\Yt_i$}
\psfrag{A}[b1]{\hspace{1em}$W$}
\psfrag{W}[]{\vspace{10em}$\Zt_i \sim \N ( 0, \a)$}
\psfrag{Z}[b1]{$Z_i \sim \N ( 0, 1)$}
\psfrag{C}[b1]{$Y_i$}
\psfrag{D}[b1]{\hspace{1em}$\Wh$}
\includegraphics[scale=0.42]{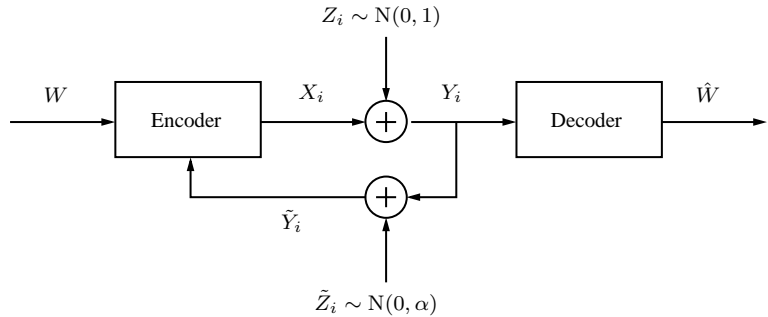}
\caption{Gaussian channel with noisy feedback.}
\end{center}
\label{fig:noisy}
\end{figure}
Suppose that the sender wishes to communicate a message $W \in [1: M] := \{1, 2, \ldots, M\}$ over the (forward) additive white Gaussian noise channel
\[
Y_i = X_i + Z_i,
\]
where $X_i$, $Y_i$, and $Z_i$ respectively denote the channel input,
channel output, and additive Gaussian noise. The sender has a causal access to a noisy version $\Yt_i$ of $Y_i$ over the feedback (backward)
additive white Gaussian noise channel
\[
\Yt_i = Y_i + \Zt_i,
\]
where $\tilde{Z_i}$ is the Gaussian noise in the backward link. We
assume that the forward noise process $\{Z_i\}_{i=1}^\infty$ and the
backward noise process $\{\Zt_i\}_{i=1}^\infty$ are independent of
each other, and respectively white Gaussian $\N(0, 1)$ and
$\N(0, \a)$.

We define an $(M, n)$ code with the encoding functions $x_i(w,
\yt^{i-1})$, $i \in [1: n]$, and the decoding function
$\wh(y^n)$. We assume a \emph{peak energy constraint}
\begin{equation}
\Ps\biggl\{\sum_{i=1}^{n}x_i^2(w, \Yt^{i-1})\leq nP\biggr\} = 1 \quad\text{for all } w\label{peak}.
\end{equation}
The probability of error of the code is defined as
\begin{align*}
P_e^{(n)}
&=\Ps\{W\ne \Wh(Y^n)\}\\
&=\frac{1}{M}\sum_{w=1}^{M}\Ps\{W\ne \Wh(Y^n)\vert W=w\},
\end{align*}
where $W$ is distributed uniformly over $\{1, 2, \ldots, M\}$ and is
independent of $(Z^n, \Zt^n)$.

As is well known, the capacity of the channel (the supremum of $(\log M)/n$ such that there exists a sequence of $(M, n)$ codes with $\lim_{n\to\infty}P_e^{(n)}\to 0$) stays the same with
or without feedback. Hence, our main focus is the reliability of communication,
which is captured by the error exponent
\[
\lim_{n\to\infty} - \frac{1}{n}\ln P_e^{(n)}
\]
of the given code. The error exponent is sensitive to the presence of noise in the
feedback link. Schalkwijk and Kailath showed in their celebrated
work \cite{Schalkwijk--Kailath1966} that \emph{noise-free} feedback can
improve the error exponent dramatically under the \emph{expected energy
constraint}
\begin{equation}
\sum_{i=1}^{n}\Es [x_i^2(w, \Yt^{i-1})]\leq nP\quad\text{for all } w\label{expected},
\end{equation}
(in fact, $P_e^{(n)}$ decays much faster than exponentially in $n$). Kim, Lapidoth, and Weissman \cite{Kim--Lapidoth--Weissman2007} studied the
optimal error exponent under the expected energy constraint
and noisy feedback, and showed that the error exponent is inversely
proportional to $\a$ for small $\a$.

Another important factor that affects the error exponent is the
energy constraint on the channel inputs---the peak energy constraint in \eqref{peak} vs.\@ the expected energy constraint in \eqref{expected}. Wyner~\cite{Wyner1968} showed that the error probability of the
Schalkwijk--Kailath coding scheme \cite{Schalkwijk--Kailath1966} degrades to
an exponential form under the peak energy constraint. In
fact, Shepp, Wolf, Wyner, and Ziv \cite{Shepp--Wolf--Wyner--Ziv1969}
showed that for the binary-message case ($M = 2$),
the best error exponent under the peak energy constraint is achieved by simple nonfeedback antipodal
signaling, regardless of the presence of feedback. This negative result might lead to an
impression that under the peak energy constraint, even noise-free
feedback does not improve the reliability of communication. Pinsker~\cite{Pinsker1968}
proved the contrary by showing that the best error exponent for
sending an $M$-ary message does not depend on $M$ and, hence can be
strictly larger than the best error exponent without feedback for $M \ge 3$.

In this paper, we show that noisy feedback can improve the
reliability of communication under the peak energy constraint, provided that
the feedback noise power $\a$ is sufficiently small. Let
\[
E_M(\a) := \limsup_{n\rightarrow\infty}-\frac{1}{n}\ln P_{e}^*(M,n),
\]
where $P_e^*(M,n)$ denotes the best error probability over all
$(M,n)$ codes for the AWGN channel with the noisy feedback.
Thus, $E_M(\infty)$ denotes the best error exponent for
communicating an $M$-ary message over the AWGN channel \emph{without} feedback.
Shannon~\cite{Shannon1959b} showed that
\begin{equation}
E_M(\infty) = \frac{M}{4(M-1)}P\label{infty}.
\end{equation}
This follows by first upper bounding the error exponent with the
sphere packing bound and then achieving this upper bound by using a
regular simplex code on the sphere of radius $\sqrt{nP}$, that is,
each codeword $x^n(w)$ satisfies $\sum_{i=1}^n x_i^2(w) =
nP$ and is at the same Euclidean distance from every other codeword. In
particular, for $M = 3$,
\begin{align*}
x^{n}(1) &= \sqrt{nP}\cdot ( 0, 1, 0, \ldots, 0),\\
x^{n}(2) &= \sqrt{nP}\cdot ( -1/2,\sqrt{3}/2, 0, \ldots, 0),\\
x^{n}(3) &= \sqrt{nP}\cdot ( -1/2,-\sqrt{3}/2, 0, \ldots, 0),
\end{align*}
and
\[
E_3(\infty) = \frac{3}{8}P.
\]
At the other extreme, $E_M(0)$ denotes the best error exponent for communicating an $M$-ary message
over the AWGN channel with \emph{noise-free} feedback.
Pinsker~\cite{Pinsker1968} showed that
\[
E_M(0) \equiv \frac{P}{2}
\]
for all $M$. In particular,
\[
E_3(0) = \frac{P}{2}.
\]
Clearly, $E_M(\a)$ is decreasing
in $\a$ and
\[
E_M(\infty) \leq E_M(\a) \leq E_M(0)
\]
for every $\a$ and $M$.

Is $E_M(\a)$ strictly larger than $E_M(\infty)$ (i.e., is noisy feedback better than no feedback)? Does $E_M(\a)$ tend to $E_M(0)$ as $\a\to 0$ (i.e., does the performance degrade gracefully with small noise in the feedback link)? What is the optimal feedback coding scheme that achieves $E_M(\a)$? To answer these questions, we establish the following results.

\begin{theorem}
For $0\le s\le 1$,
\[
E_M(\a^*(s)) \ge \frac{P}{2}\bl 1 - \frac{3(M - 2)}{M(s^2 - 2s + 4) + 3(M - 2)}\br,
\]
where
\[
\a^*(s) = \frac{3s^2}{4(s^2 - 2s + 4)}.
\]
\end{theorem}

By comparing the lower bound with \eqref{infty} and identifying the critical point $\a = \a^*(1) = 1/4$, we obtain the following.
\begin{corollary}
\[
E_M(\a)>E_M(\infty)\quad\text{for } \a < \frac{1}{4}.
\]
\end{corollary}

Thus, if the noise power in the feedback link is sufficiently small, then
the noisy feedback improves the reliability of communication even
under the peak energy constraint. The proof of Theorem~$1$ is motivated by recent
results of Burnashev and Yamamoto in a series of
papers~\cite{Burnashev--Yamamoto2008a}, \cite{Burnashev--Yamamoto2008b}, where they considered
a communication model with a forward BSC($p$) and a
backward BSC($\a p$), and showed that when $\a$ is sufficiently small, the best error
exponent is strictly larger than the one without feedback.

The lower bound in Theroem~$1$ shows that $\liminf_{\a\to 0}E_M(\a)\ge 2PM/(7M - 6)$, which is strictly less than $E_M(0) = P/2$. To obtain a better asymptotic behavior for $\a\to 0$, we establish the following.
\begin{theorem}
\begin{align*}
 E_M(\alpha)
&\ge\frac{P}{2}\frac{1}{1 + \a + 4(\lfloor M/2 \rfloor)^2 \a + 4(\lfloor M/2 \rfloor)\sqrt{\a(1 + \a)}}\\
&\ge \frac{P}{2}\frac{1}{(\sqrt{\a}M + \sqrt{1 + \a})^2}.
\end{align*}
\end{theorem}

This theorem leads to the following.
\begin{corollary}
\[
\lim_{\a\to 0}E_M(\a) = E_M(0).
\]
\end{corollary}

Thus, the lower bound in Theorem~$2$ is tight for $\a \to 0$. The proof of Theorem~$2$ extends Pinsker's linear noise-free feedback coding scheme~\cite{Pinsker1968} to the noisy case.

\begin{figure}[htbp]
\centering
\footnotesize
\psfrag{C}[b]{${E_3'(\alpha)}$}
\psfrag{A}[b]{${E_3(0)}$}
\psfrag{B}[b]{${E_3''(\alpha)}$}
\psfrag{D}[b]{${E_3(\infty)}$}
\psfrag{M}[b]{$\a$}
\psfrag{N}[b]{$E_3(\a)$}
\psfrag{E}[b]{$5.6\times 10^{-3}$}
\psfrag{F}[b]{$0.25$}
\hspace{-6pt}\includegraphics[scale=0.5]{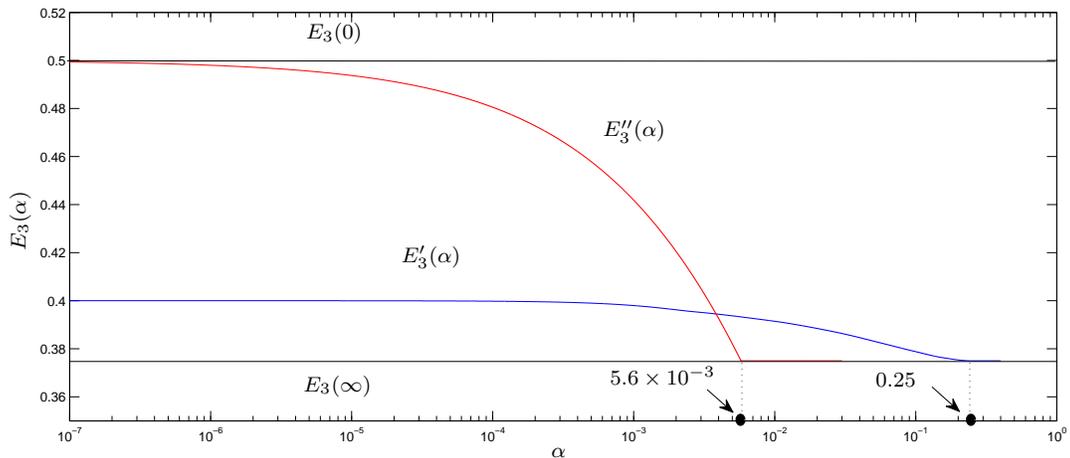}
\caption{Comparison of the two noisy feedback coding scheme for $M = 3$.}
\label{fig:compare}
\end{figure}

Fig.~\ref{fig:compare} compares the two bounds for the $M = 3$ case. The linear noisy feedback coding scheme performs better when $\a$ is sufficiently small, while the two-stage noisy feedback coding scheme performs better when $\a$ is relatively larger.

The rest of the paper is organized as follows. In Section II, we study a two-stage noisy feedback coding scheme motivated by recent results of Burnashev and Yamamoto and establish Thereom~$1$. In Section III, we extends Pinsker's noise-free linear feedback coding scheme to the noisy feedback case and establish Theorem~$2$. Section~IV concludes the paper.

\section{Two-stage Noisy Feedback Coding Scheme}

\subsection{Background}
It is instructive to first consider a two-stage noise-free feedback coding scheme for $M = 3$.
This two-stage scheme has been studied by Schalkwijk and Barron~\cite{Schalkwijk--Barron1971} and
Yamamoto and Itoh~\cite{Yamamoto--Itoh1979} for a general $M$.

\smallskip\noindent{\bf Encoding.}
Fix some $\l \in (0, 1)$. For simplicity of notation, assume throughout that $\l n$ is an integer.
To send message $w\in [1: 3]$, during the transmission time
interval $[1: \l n]$ (namely, stage $1$), the encoder uses the simplex signaling:
\begin{equation}
x^{\l n}(w) =
\begin{cases}
\sqrt{\l nP}\cdot ( 0, 1, 0, \ldots, 0) & \text{ for }w = 1,\\
\sqrt{\l nP}\cdot ( -1/2,\sqrt{3}/2, 0, \ldots, 0) & \text{ for }w = 2,\\
\sqrt{\l nP}\cdot ( -1/2,-\sqrt{3}/2, 0, \ldots, 0) & \text{ for }w = 3.\label{two_stage}
\end{cases}
\end{equation}
Based on the feedback $y^{\l n}$, the encoder then chooses the two most probable message estimates $\wh_1$ and $\wh_2$, where
\begin{equation}
p(\wh_1|y^{\l n})\ge p(\wh_2|y^{\l n}) \ge p(\wh_3|y^{\l n})\label{order}
\end{equation}
and in case of a tie the one with the smaller index is chosen. Since the channel is Gaussian and $W$ is uniform, $\eqref{order}$ can be written as
\[
||x^{\l n}(\wh_1) - y^{\l n}||\ge ||x^{\l n}(\wh_2) - y^{\l n}||\ge ||x^{\l n}(\wh_3) - y^{\l n}||,
\]
where $||\cdot||$ denotes the Euclidean distance.
During the transmission time interval $[\l n+1: n]$ (stage $2$), the encoder uses antipodal signaling for $w$ if $w \in \{\wh_1, \wh_2\}$
and transmits all-zero sequence otherwise:
\begin{equation*}
x^{n}_{\l n + 1}(w) =
\begin{cases}
\sqrt{(1 - \l) nP}\cdot ( 1, 0, 0, \ldots, 0) & \text{ if } w = \min\{\wh_1, \wh_2\},\\
\sqrt{(1 - \l) nP}\cdot ( -1, 0, 0, \ldots, 0) & \text{ if } w = \max\{\wh_1, \wh_2\},\\
( 0, 0, \ldots, 0) & \text{ otherwise. }
\end{cases}
\end{equation*}

\smallskip\noindent{\bf Decoding.}
At the end of stage $1$, the decoder chooses the two most probable message estimates $\wh_1$ and $\wh_2$ based on $Y^{\l n}$ as the encoder does.
At the end of stage $2$, the decoder declares that $\wh$ is sent if
\begin{align*}
\wh
&= \arg\min_{w \in \{\wh_1, \wh_2\}}||x^n(w) - y^n|| \\
&= \arg\min_{w \in \{\wh_1, \wh_2\}} \bigl( ||x^{\l n}(w) - y^{\l n}||^2 + ||x^{n}_{\l n + 1}(w) - y_{\l n + 1}^n||^2\bigr)^{1/2}.
\end{align*}

\begin{figure}[htbp]
\centering
{\footnotesize\rmfamily
\psfrag{1}[b]{\hspace{2em}$1$} \psfrag{2}[b]{$2$} \psfrag{3}[b]{\hspace{.3em}$3$} \psfrag{A}[b]{\hspace{-.3em}$A_{12}$}
\psfrag{B}[b]{$A_{13}$} \psfrag{C}[b]{\hspace{1em}$A_{23}$}
\psfrag{d}[b]{$d_1$}
\includegraphics[scale=0.3]{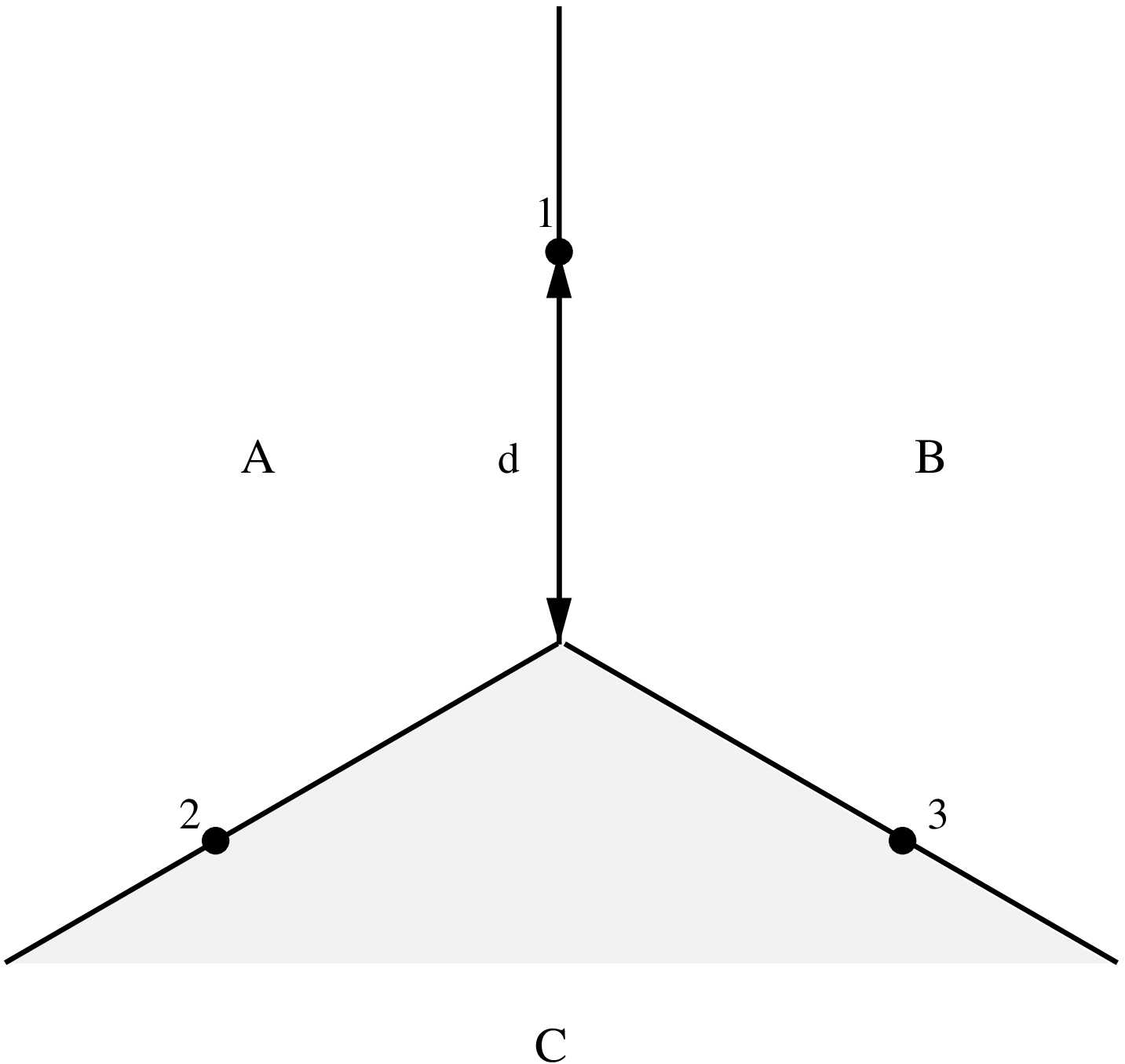}}
\caption{The error event $\Ec_1$ when $W = 1$. Here $d_1 = \sqrt{\l nP}$ and $1$, $2$, and $3$
denote $x^{\l n}(1)$, $x^{\l n}(2)$, and $x^{\l n}(3)$, respectively.}
\label{fig:error1}
\end{figure}

\smallskip\noindent{\bf Analysis of the probability of error.}
Let $\Wh_1$ and $\Wh_2$ denote the two most probable message estimates at the end of stage $1$.
The decoder makes an error if and only if one of the following events occurs:
\begin{align*}
\Ec_1 &= \blb W \ne \Wh_1 \text{ and } W \ne \Wh_2\brb,\\
\Ec_2 &= \blb W\in\{\Wh_1, \Wh_2\}\text{ and } \Wh \ne W\brb.
\end{align*}
Thus, the probability of error is
\[
P_e^{(n)} = \P(\Ec_1)  + \P(\Ec_2).
\]
By symmetry, we assume without loss of generality that $W = 1$ is sent. For brevity, we do not explicitly condition on the event $\{W = 1\}$
in probability expressions in the following, whenever it is clear from the context. Refering to Fig.~\ref{fig:error1}, let
\[
A_{23} = \bigl\{y^{\l n}\suchthat ||x^{\l n}(1) - y^{\l n}||\ge ||x^{\l n}(2) - y^{\l n}||\text{ and }
||x^{\l n}(1) - y^{\l n} ||\ge ||x^{\l n}(3) - y^{\l n}||\bigr\},
\]
we have
\begin{align*}
\P(\Ec_1)
&= \P\{Y^{\l n}\in A_{23}\} \\
&\le Q(d_1)\\
&\overset{(a)}{\le}\frac{1}{2}\exp\bl-\frac{\l nP}{2}\br,
\end{align*}
where $(a)$ follows since $Q(x) \le (1/2)\exp(-x^2/2) \text{ for } x \ge 0$ (see \cite[Problem $2.26$]{Wozencraft--Jacobs1965}).

\begin{figure}[htbp]
\centering
{\footnotesize\rmfamily
\psfrag{1}{\hspace{-.1em}$1$}
\psfrag{2}{\hspace{-.5em}$2$}
\psfrag{3}{$3$}
\psfrag{d}{\hspace{-1em}$d_2$}
\includegraphics[scale=0.3]{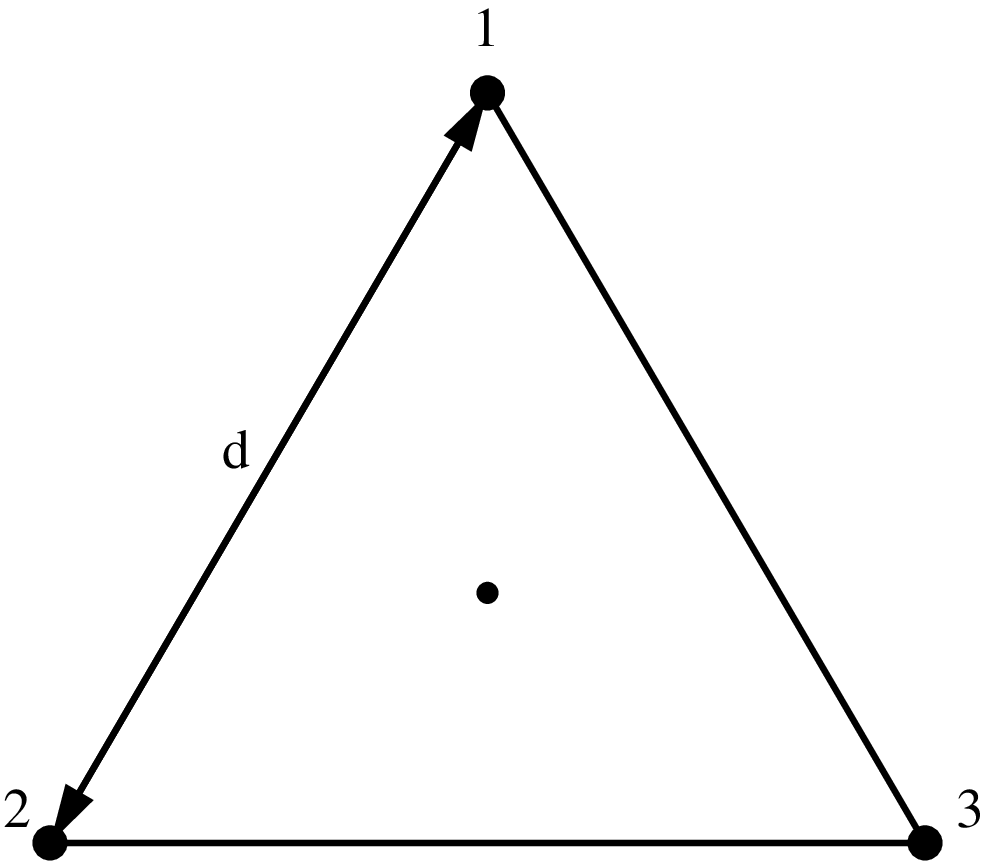}
\hspace{4em}
\psfrag{a}{\hspace{-.5em}$\wh_1$}
\psfrag{b}{$\wh_2$}
\psfrag{d}{\hspace{-.5em}$d_3$}
\includegraphics[scale=0.3]{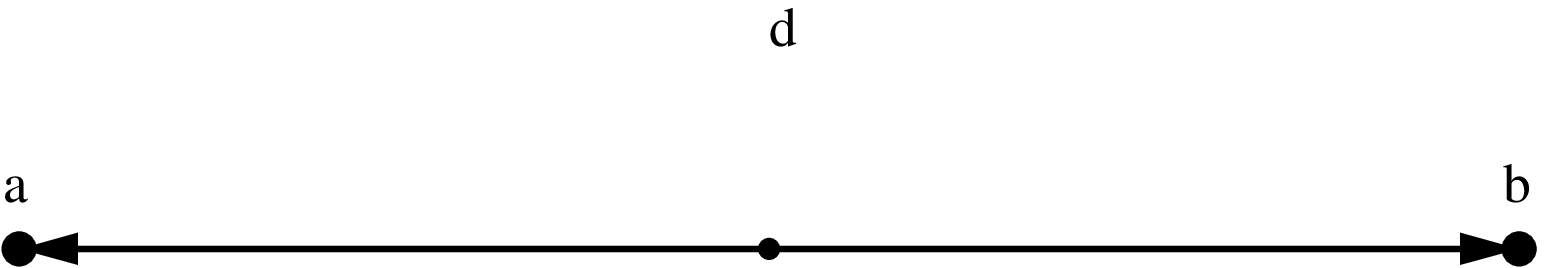}
}
\caption{The error event $\Ec_2$. Here $d_2 = \sqrt{3\l nP}$
and $d_3 = \sqrt{4(1-\l)nP}$.}
\label{fig:error2}
\end{figure}

On the other hand, $\P(\Ec_2)$ is determined by the distance between the
simplex signaling in stage $1$ and
the distance between the antipodal signaling in stage $2$ (see Fig.~\ref{fig:error2}).
In particular,
\[
||X^{n}(\Wh_1) - X^{n}(\Wh_2)|| = \sqrt{d_2^2 + d_3^2} = \sqrt{(4 - \l)nP}.
\]
Thus,
\begin{align*}
\P(\Ec_2)
&=Q\bl\frac{||X^n(\Wh_1) - X^n(\Wh_2)||}{2}\br\\
&= Q\bl\sqrt{\bl 1 - \frac{\l}{4}\br nP}\br\\
&\le \frac{1}{2}\exp\bl -\frac{1}{2}\bl 1 - \frac{\l}{4}\br nP\br.
\end{align*}
Therefore, the error exponent of the two-stage feedback coding scheme is lower bounded as
\begin{align*}
E_3'(0)
&=\limsup_{n\to \infty}-\frac{1}{n}\ln P_e^{(n)}\\
&=\limsup_{n\to\infty}-\frac{1}{n}\max\{\ln \P(\Ec_1), \ln \P(\Ec_2)\}\\
&\ge \min\biggl\{\frac{\l P}{2}, \frac{P}{2}\bl 1 - \frac{\l}{4}\br \biggr\}.
\end{align*}
Now let $\l = 4/5$. Then it can be readily verified that both terms in the minimum are the same and we have
\[
E_3(0) \ge E_3'(0) \ge \frac{2P}{5}.
\]

\begin{remark}
Since $E_3(0) = P/2$, this two-stage noise-free feedback coding scheme is strictly suboptimal.
\end{remark}
\begin{remark}
We need only three transmissions: two for stage $1$ and one for stage $2$. Thus $\l$ actually divides only the total energy $nP$, not the block length $n$.
\end{remark}
\subsection{Two-stage Noisy Feedback Coding Scheme}
Based on the two-stage noise-free feedback coding scheme in the previous subsection and a new idea of {\em signal protection} introduced by
Burnashev and Yamamoto~\cite{Burnashev--Yamamoto2008a}, \cite{Burnashev--Yamamoto2008b},
we present a two-stage noisy feedback coding scheme for $M = 3$. The coding scheme for an arbitrary $M$ is given in Appendix~A.

\begin{figure}[htbp]
\centering
{\footnotesize\rmfamily
\psfrag{A}[b]{$A_{12}'$}\psfrag{B}[b]{\hspace{1em}$A_{13}'$} \psfrag{C}[b]{\hspace{1em}$A_{23}'$}
\psfrag{D}[b]{\hspace{1em}$B_1$} \psfrag{E}[b]{$B_2$} \psfrag{F}[b]{$B_3$}
\psfrag{1}[b]{\hspace{.3em}$1$} \psfrag{2}[b]{$2$} \psfrag{3}[b]{\hspace{0.3em}$3$}
\psfrag{s}[b]{\hspace{1em}$d_1$}  \psfrag{t}[b]{$d_4$}
\psfrag{z}[b]{$d'$}
\includegraphics[scale=0.4]{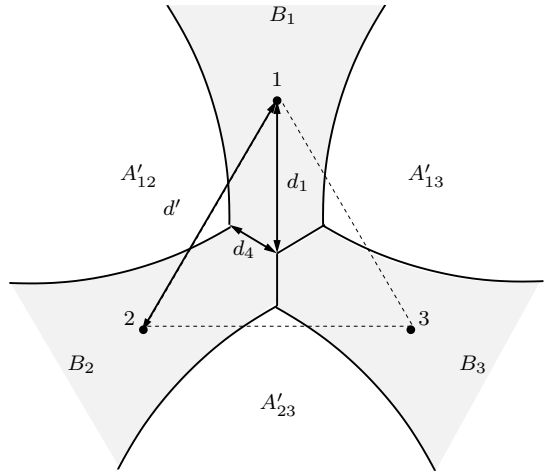}}
\caption{Signal protection regions. The shaded areas $B_w$ for $w = 1, 2, 3$ are the signal protection
regions for $x^{\l n}(1)$, $x^{\l n}(2)$, and $x^{\l n}(3)$, respectively. Here $d_4 = s d_1/2 = (s/2)\sqrt{\l nP}$ for some parameter $s = s(t)\in [0, 1]$ to be optimized later. }
\label{fig:decision}
\end{figure}

In the two-stage noise-free feedback coding scheme, the encoder and decoder agree on the same set of message estimates $\wh_1$ and $\wh_2$ at the end of stage $1$.
When there is noise in the feedback link, however, this coordination is not always possible.
To solve this problem, we assign a signal protection region $B_w$, $w\in[1: 3]$, to each signal $x^{\l n}(w)$ as depicted in Fig.~\ref{fig:decision}.
Let $x^{\l n}$ and $y^{\l n}$ denote the transmitted and
received signals, respectively, and $\tilde{y}^{\l n}$ denote the feedback sequence at the encoder.
Let $d' =||x^{\l n}(1) - x^{\l n}(2)|| = \sqrt{3\l nP}$ and the signal protection region $B_w$ for $x^{\l n}(w)$, $w\in [1: 3]$, is defined as
\begin{align}
B_w = \bigl\{y^{\l n}\suchthat & ||x^{\l n}(w) - y^{\l n}|| \le ||x^{\l n}(w') - y^{\l n}||\text{ for }w'\ne w\text{ and }\nonumber\\
        & \bigl|||x^{\l n}(w') - y^{\l n}|| - ||x^{\l n}(w'') - y^{\l n}||\bigr|\le td'\text{ for }w', w''\ne w\bigr\}\label{protection}
\end{align}
which means that message $w$ is the most probable and the other messages $w'$ and $w''$ are of approximately equal posterior probabilities. Here $t\in [0, (\sqrt{3} - 1)/2]$ is a fixed parameter which will be optimized later in the analysis.

\smallskip\noindent{\bf Encoding.}
In stage $1$, the encoder uses the same simplex signaling as in the noise-free feedback case (see \eqref{two_stage}).
Then based on the noisy feedback $\yt^{\l n}$, the encoder chooses $\wt_1$ and $\wt_2$ such that
\[
||x^{\l n}(\wh_1) - y^{\l n}||\ge ||x^{\l n}(\wh_2) - y^{\l n}||\ge ||x^{\l n}(\wh_3) - y^{\l n}||,
\]
In stage $2$, the encoder uses antipodal signaling for $w$ if $w\in\{\wt_1, \wt_2\}$ and transmits all-zero sequence otherwise.

\smallskip\noindent{\bf Decoding.}
The decoder makes a decision immediately at the end of stage $1$ if the received signal lies in one of the signal protection regions, i.e., $y^{\l n}\in B_w$ for $w \in [1: 3]$.
Otherwise, it chooses the two most probable message estimates $\wh_1$ and $\wh_2$ and wait for the transmission in stage $2$.
At the end of stage $2$, the decoder declares that $\wh$ is sent if
\begin{align*}
\wh
&= \arg\min_{w \in \{\wh_1, \wh_2\}}||x^n(w) - y^n|| \\
&= \arg\min_{w \in \{\wh_1, \wh_2\}} \bigl( ||x^{\l n}(w) - y^{\l n}||^2 + ||x^{n}_{\l n + 1}(w) - y_{\l n + 1}^n||^2\bigr)^{1/2}.
\end{align*}

\begin{remark}
The signal protection region corresponds to the case in which the two least probable messages are of approximately equal posterior probabilities,
i.e.,
$||x^{\l n}(w) - y^{\l n}||\ll ||x^{\l n}(w') - y^{\l n}||\approx ||x^{\l n}(w'') - y^{\l n}||$.
\end{remark}

\smallskip\noindent{\bf Analysis of the probability of error.}
Let ($\Wt_1$, $\Wt_2$) and ($\Wh_1$, $\Wh_2$) denote the pairs of the two most probable message estimates at the encoder and the decoder, respectively. As before, we assume that $W = 1$ is sent. Refering to Fig.~\ref{fig:decision}, let
\[
A_{ww'}' = A_{ww'}\backslash(\cup_{w''}B_{w''}),\quad w, w'\in [1: 3]
\]
where $
A_{ww'} = \{y^{\l n}\suchthat ||y^{\l n} - x^{\l n}(w'')||\ge \max\{||y^{\l n} - x^{\l n}(w)||, ||y^{\l n} - x^{\l n}(w')||\},\; w''\ne w, w' \}$.

\begin{figure}[t]
\hspace{-20em}\centering
{\psfrag{x}[b]{$Y^{\l n}$}
\psfrag{A}[b]{$A_{12}'$}
\psfrag{B}[b]{$A_{13}'$}
\psfrag{C}[b]{\hspace{1em}$A_{23}'$}
\psfrag{D}[b]{\hspace{1em}$B_1$} \psfrag{E}[b]{$B_2$} \psfrag{F}[b]{$B_3$}
\psfrag{1}[b]{\hspace{.3em}$1$}
\psfrag{2}[b]{$2$}\psfrag{3}[b]{\hspace{0.3em}$3$}
\psfrag{s}[b]{\hspace{1em}$d_1$}
\psfrag{t}[b]{\vspace{0.5em}$d_4$}
\psfrag{k}[b]{\hspace{0.15em}$d_5$}
\includegraphics[scale=0.40]{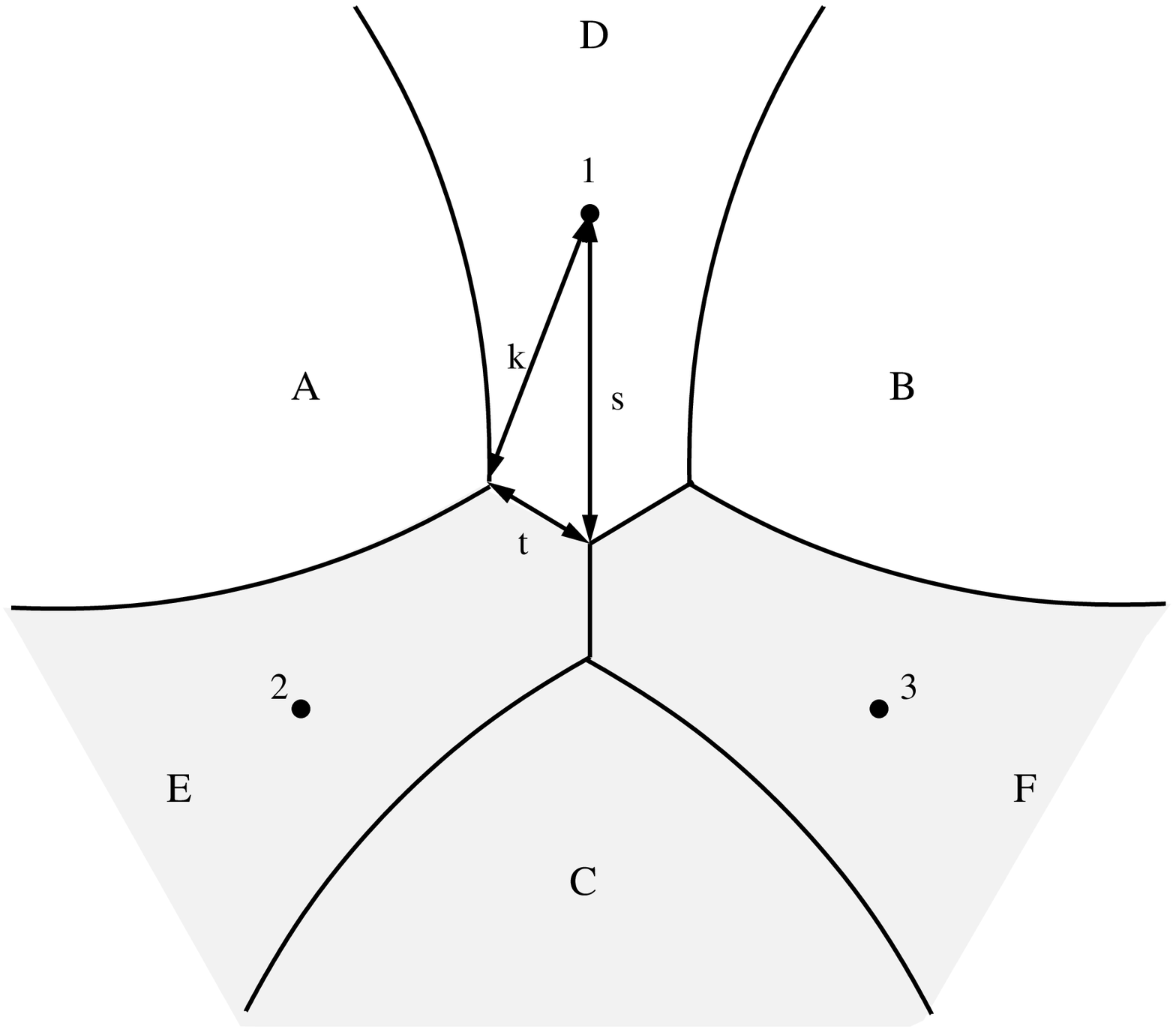}\\
\hspace{-20em}{\footnotesize (a)}\\[3em]
}
\vspace{-26.3em}
\hspace{20em}\centering
{\psfrag{x}[b]{$Y^{\l n}$}
\psfrag{A}[b]{$A_{12}'$}\psfrag{B}[b]{$A_{13}$} \psfrag{C}[b]{\hspace{1em}$A_{23}$}
\psfrag{1}[b]{$1$}
\psfrag{2}[b]{$2$}\psfrag{3}[b]{\hspace{.3em}$3$}
\psfrag{t}[b]{\hspace{.5em}$d_4$}
\psfrag{s}[b]{$d_6$}
\includegraphics[scale=0.40]{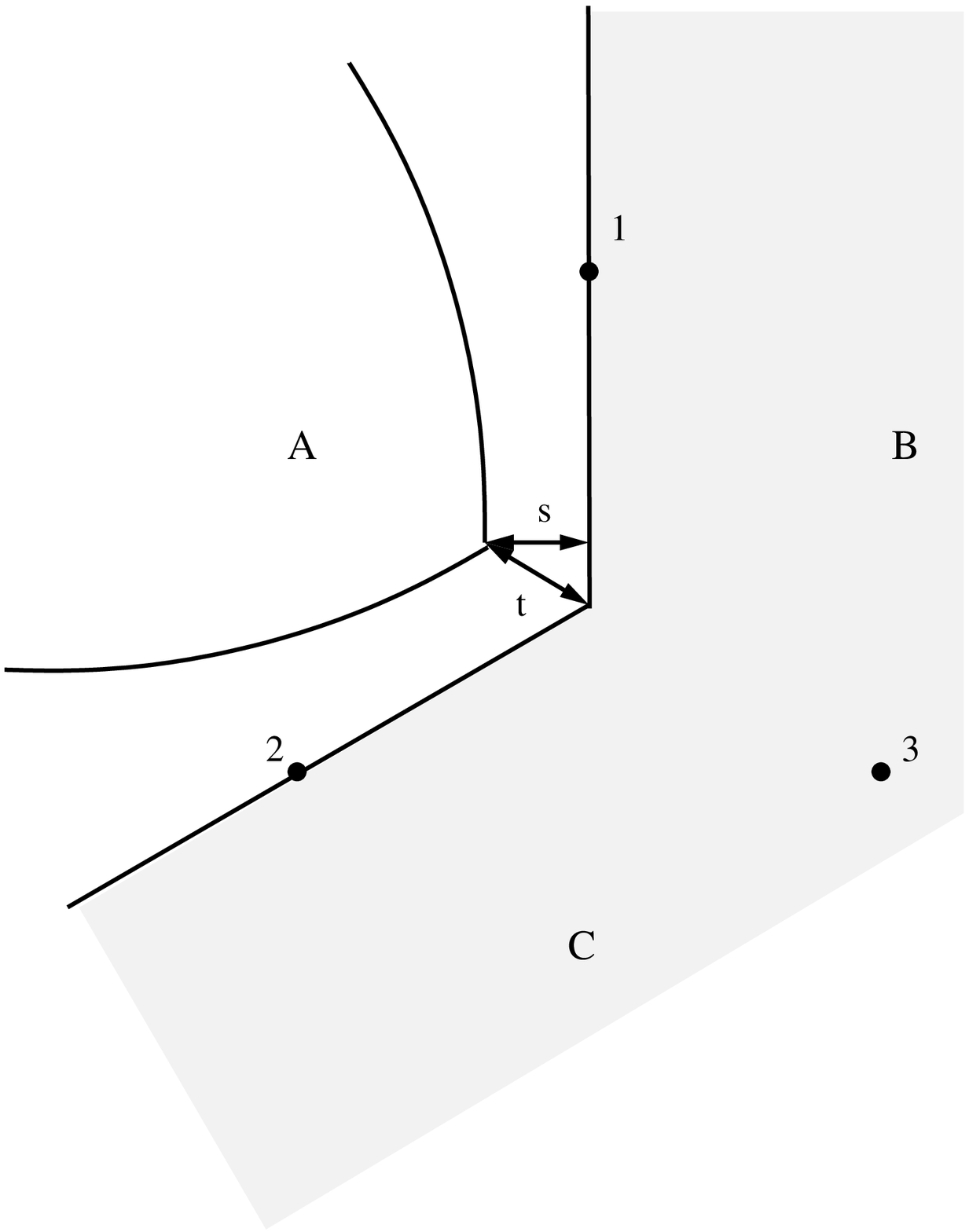}\\
\hspace{20em}{\footnotesize (b) }
}
\caption{(a) The error event $\Ec_1$ when $W = 1$. Since $0\le s\le 1$, we have $d_5 = \sqrt{d_1^2 + d_4^2 - d_1d_4} = \sqrt{(\l nP/4)(s^2 - 2s + 4)} $.
(b) The error event $\Ec_2$ when $W = 1$ and $\{\Wh_1, \Wh_2\} = \{1, 3\}$. Here $d_6 = (\sqrt{3}/2)d_4 = s\sqrt{(3\l n P/16)}$.}
\label{fig:error_events}
\end{figure}

The decoder makes an error only if one or more of the following events occur:

\begin{itemize}
\item decoding error at the end of stage $1$
\[
\Ec_1 = \blb Y^{\l n}\in B_2\cup B_3\cup A_{23}'\brb,
\]
\item miscoordination due to the feedback noise
\begin{align*}
\tilde{\Ec}_{12} &= \blb Y^{\l n}\in A_{12}', \Yt^{\l n}\in A_{13}\cup A_{23}\brb,\\
\tilde{\Ec}_{13} &= \blb Y^{\l n}\in A_{13}', \Yt^{\l n}\in A_{12}\cup A_{23}\brb,
\end{align*}
\item decoding error at the end of stage $2$
\[
\Ec_2 = \blb W\in\{\Wh_1, \Wh_2\} = \{\Wt_1, \Wt_2\} \text{ and }\Wh \ne W\brb.
\]
\end{itemize}
Thus, the probability of error is upper bounded as
\begin{align*}
P_e^{(n)} &\le \P(\Ec_1) + \P(\tilde{\Ec}_{12}) + \P(\tilde{\Ec}_{13}) + \P(\Ec_2)\\
&= \P(\Ec_1) + 2\P(\tilde{\Ec}_{12}) + \P(\Ec_2).
\end{align*}
To simplify the analysis, we introduce
a new parameter $s\in [0, 1]$ such that $d_4 = s d_1/2 = (s/2)\sqrt{\l nP}$. It can be easily checked that $s\in [0, 1]$ corresponds to
$t\in [0, (\sqrt{3} - 1)/2]$ and that this constraint guarantees that $d_5 = \min_{y^{\l n}\in A_{23}'\cup B_2\cup B_3}||x^{\l n}(1) - y^{\l n}||$
(see Fig.~\ref{fig:error_events}(a)).
Hence, for the first term
\begin{align}
\P(\Ec_1)
&=\P\blb Y^{\l n} \in A_{23}'\cup B_2\cup B_3\brb\nonumber\\
&\le 2Q(d_5)\label{d_5}\\
&\le \exp\bl-\frac{\l nP}{8}(s^2 - 2s + 4)\br.\nonumber
\end{align}
The second term $\P(\tilde{\Ec}_2)$ can be upper bounded (see Fig.~\ref{fig:error_events}(b)) as
\begin{align}
\P(\tilde{\Ec}_{12})
&= \P\blb Y^{\l n}\in A_{12}', \Yt^{\l n}\in A_{13}\cup A_{23}\brb\nonumber\\
&\le \P\blb\Yt^{\l n}\in A_{13}\cup A_{23} | Y^{\l n}\in A_{12}'\brb\nonumber\\
&\le 2Q\bl \frac{d_6}{\sqrt{\a}}\br\label{d_6}\\
&\le \exp\bl-\frac{3s^2\l nP}{32\a}\br.\nonumber
\end{align}
\noindent Finally, the third term $\P(\Ec_3)$ can be upper bounded in the exactly same manner as in the noise-free feedback case:
\[
\P(\Ec_3)\le \frac{1}{2}\exp\bl -\frac{1}{2}\bl 1 - \frac{\l}{4}\br nP\br.
\]
Therefore, the error exponent of the two-stage noisy feedback coding scheme is lower bounded as
\begin{align*}
E_3'(\alpha)
&= \limsup_{n\to\infty}- \frac{1}{n}\ln P_e^{(n)}\\
&\ge \limsup_{n\to\infty} -\frac{1}{n} \max\{\ln \P(\Ec_1), \ln \P(\tilde{\Ec}_{12}), \ln \P(\Ec_2)\}\\
&\ge \min \biggl\{\frac{\l P}{8}(s^2 - 2s + 4),\;
\frac{3\l s^2 P}{32\a},\;
\frac{P}{2}\bl 1 - \frac{1}{4}\l\br\biggr\}.
\end{align*}
Now let
\[
\a = \a^*(s) = \frac{3s^2}{4(s^2 - 2s + 4)}
\]
and
\[
\l = \l^*(s) = \frac{4}{s^2 - 2s +5}.
\]
Then it can be readily verified that all the three terms in the minimum are the same and we have
\begin{equation}
E_3'(\a^*(s)) \ge \frac{P}{2}\frac{s^2 - 2s + 4}{s^2 - 2s + 5}=: \phi(s)\label{E_3}.
\end{equation}
Note that if $s < 1$,
\[
\phi(s)> \frac{3}{8}P = E_3(\infty),
\]
and $\a^*(s)$ is
monotonically increasing over $s\in [0, 1]$. Thus
\[
E_3(\a) > E_3(\infty)\quad\text{for } \a < \a^*(1) = \frac{1}{4}.
\]
This completes the proof of Theorem $1$ for the $M = 3$ case.
\begin{remark}
It can be easily checked that the lower bound in $\eqref{E_3}$ is tight and characterizes the exact error exponent $E_3'(\a)$ of the two-stage noisy feedback coding scheme.
\end{remark}

\section{Linear Noisy Feedback Coding Scheme}

\subsection{Background}
It is instructive to revisit (a slightly simplified version of) the linear noise-free feedback coding scheme by Pinsker~\cite{Pinsker1968}, which shows that $E_M(0) \ge E_2(\infty) = P/2$ for all $M \ge 2$. This lower
bound is tight since $E_2(0) = E_2(\infty)$~\cite{Shepp--Wolf--Wyner--Ziv1969} and $E_M(0)$ is nonincreasing in $M$.

\smallskip\noindent{\bf Encoding.}
To send message $w\in [1: M]$, the encoder transmits
\begin{equation}
X_1(w) =
\begin{cases}
\frac{L + 1 - w}{L}\sqrt{P} & \text{ if }M = 2L + 1,\\
\frac{L + 1/2 - w}{L}\sqrt{P} & \text{ if }M = 2L.\label{X_1}
\end{cases}
\end{equation}
Because of the feedback $Y_1$, the encoder can learn the noise $Z_1 = Y_1 - X_1$. Subsequently it transmits
\[
X_i = (1 + \d)Z_{i-1},\quad i \in[2: \eta],
\]
and $X_i = 0$ afterwards, where $\d > 0$ will be optimized later and the random time $\eta = \eta(w, Z^n)$ is the largest $k \le \nb = \sqrt{n}$ such that
\[
 \sum_{i=1}^{k}X_i^2 \le nP.
\]

\smallskip\noindent{\bf Decoding.}
Upon receiving $Y^n$, the decoder estimates $X_1$ by
\[
\Xh_1 = \sum_{i=1}^{\nb}(-1)^{i-1}\frac{Y_i}{(1+\d)^{i-1}}
\]
and declares that $\wh$ is sent if
\[
\wh = \arg\min_{w\in [1: M]}|X_1(w) - \Xh_1|.
 \]
\begin{remark}
It can be easily checked that each time $i\in [2: \eta]$, the encoder transmits the error
\[
\sum_{j = 1}^{i - 1}(-1)^{j - 1}\frac{Y_j}{(1 + \d)^{j - 1}} - X_1 = (-1)^{i - 2}\frac{Z_{i - 1}}{(1 + \d)^{i - 2}}
\]
in the decoder's current estimate of the initial transmission (up to scaling). Thus, Pinsker's coding scheme is another instance of iterative refinement used in the Schalkwijk-Kailth coding scheme~\cite{Schalkwijk--Kailath1966} for the Gaussian channel and the Horstein coding scheme~\cite{Horstein1963} for the binary symmetric channel.
\end{remark}

\smallskip\noindent{\bf Analysis of the probability of error.}
For simplicity of notation, assume throughout that $\nb = \sqrt{n}$ is an integer. We use $\eps_n$ to denote a generic sequence of nonnegative numbers that tends to zero
as $n\to \infty$. When there are multiple such functions $\eps_n^{(1)}, \eps_n^{(2)},\cdots, \eps_n^{(k)}$, we denote them all by $\eps_n$ with the understanding
that $\eps_n = \max\{\eps_n^{(1)}, \eps_n^{(2)},\cdots, \eps_n^{(k)}\}$.
It is easy to see that decoding error occurs only if $|X_1(w) - \Xh_1| > \sqrt{P}/(2L)$. The probability of error is thus upper bounded as
\[
P_{\error}^{(n)} = \P\{W \ne \Wh\} \le \P\biggl\{|X_1 - \Xh_1| > \frac{\sqrt{P}}{2L}\biggr\}.
\]

The key idea in the analysis is to introduce a ``virtual'' transmission
\begin{equation}
X_i' = \begin{cases}
X_1  & \textrm{if $i = 1$},\\
(1 + \d)Z_{i-1}&\textrm{if $i \in [2: \nb]$},\\
0  &\textrm{otherwise}.
\end{cases}
\end{equation}
Let
\begin{equation}
Y_i' = X_i' + Z_i \label{Y_i'}
\end{equation}
and define the estimate $\Xh_1'$ of $X_1'$ as

\begin{equation}
\Xh_1' = \sum_{i=1}^{\nb}(-1)^{i-1}\frac{Y_i'}{(1+\d)^{i-1}}. \label{Xh_1'}
\end{equation}
Then, it can be easily shown that
\[
\Xh_1' = X_1 + (-1)^{\nb - 1}\frac{Z_{\nb}}{(1 + \d)^{\nb - 1}}.
\]
Thus we have
\begin{align*}
\P\biggl\{|X_1 - \Xh_1| > \frac{\sqrt{P}}{2L}\biggr\} &\le \P\biggl\{|X_1 - \Xh_1'| + |\Xh_1' - \Xh_1| > \frac{\sqrt{P}}{2L}\biggr\}\\
&\le \P\biggl\{|X_1 - \Xh_1'| > \frac{\sqrt{P}}{2L}\biggr\} + \P\{|\Xh_1' - \Xh_1| > 0\}\\
&=: P_1 + P_2.
\end{align*}

Now we upper bound the two terms. For the first term, we have
\begin{align*}
P_1
&= \P\biggl\{\left|\frac{Z_{\nb}}{(1 + \d)^{\nb - 1}}\right| > \frac{\sqrt{P}}{2L}\biggr\}\\
&=2Q\bl\frac{\sqrt{P}(1+\d)^{\nb - 1}}{2L}\br\\
&\le\exp\bl-\frac{P(1 + \d)^{2(\nb - 1)}}{8L^2}\br.
\end{align*}
For the second term, note that $X_i = X_i'$ for all $i$ if and only if $\sum_{i=1}^{\nb}X_i^2 \le nP$, and thus that $\Xh_1' \ne \Xh_1$ only if $\sum_{i=1}^{\nb}X_i^2 > nP$. Therefore,
\begin{align*}
P_2
&\le \P\biggl\{\sum_{i=1}^{\nb}X_i^2 > nP\biggr\}\\
& \overset{(a)}{\le} \P\biggl\{\sum_{i=2}^{\nb}(1+\d)^2 Z_{i-1}^2> (n-1)P\biggr\}\\
&= \P\biggl\{\chi_{\nb - 1}^2 > \frac{(n - 1)P}{(1 + \d)^2}\biggr\},
\end{align*}
where $(a)$ follows since $X_1^2 \le P$ (recall $\eqref{X_1}$) and $\chi^2_{\nb - 1}$ denotes a chi-square random variable with $\nb - 1$
degrees of freedom.
By upper bounding the tail probability of the chi-square random variable~\cite{Inglot--Ledwina2006} as

\begin{equation}
\P\{\chi_k^2 > x\} \le \exp\bl -\frac{x}{2} + \frac{k}{2}\log\frac{ex}{k} \br \quad\text{ for any } k \ge 1 \text{ and } x \ge k, \label{Chi}
\end{equation}
we have
\begin{align*}
P_2
&\le \P\biggl\{\chi_{\nb - 1}^2 > \frac{(n - 1)P}{(1 + \d)^2}\biggr\}\\
&\le \exp\bl -\frac{1}{2}\frac{(n - 1)P}{(1 + \d)^2} + \frac{\nb - 1}{2}\log\frac{e(n - 1)P}{(\nb - 1)(1 + \d)^2}\br\\
&\le \exp\bl -\frac{1}{2}\frac{(n - 1)P}{(1 + \d)^2} + \frac{\nb - 1}{2}\log\frac{e(n - 1)P}{(\nb - 1)}\br\\
&\le\exp\bl -\frac{1}{2}\frac{nP}{(1 + \d)^2} + n\eps_n\br,
\end{align*}
where $\eps_n$ tends to zero as $n\to\infty$.
Therefore, the error exponent of the linear feedback coding scheme is lower bounded as
\begin{align*}
E_M''(0)
&\ge \limsup_{n\to\infty}- \frac{1}{n}\ln P_e^{(n)}\\
&= \limsup_{n\to\infty} -\frac{1}{n} \max\{\ln P_1, \ln P_2\}\\
&\ge \limsup_{n\to\infty}\min \biggl\{\frac{P(1 + \d)^{2(\nb - 1)}}{8nL^2},\; \frac{P}{2(1 + \d)^2} \biggr\}.
\end{align*}
for any $\d > 0$. Now let
\[
\d = \d(n) = \frac{\ln(4nL^2)}{2\nb},
\]
which tends to zero as $n\to\infty$. Then the limits of both terms in the minimum are the same. Therefore,
\[
E_M''(0) \ge \limsup_{n\to\infty} \frac{P}{2(1 + \d(n))^2} = \frac{P}{2},
\]
which completes the proof of achievability.

\subsection{Linear Noisy Feedback Coding Scheme}
Now we formally describe and analyze a linear noisy feedback coding scheme based on Pinsker's noise-free feedback coding scheme.

\smallskip\noindent{\bf Encoding.}
Fix some $\l \in (0, 1)$. To send message $w\in [1: M]$, the encoder transmits
\begin{equation}
X_1(w) =
\begin{cases}
\frac{L + 1 - w}{L}\sqrt{\l nP} & \text{ if }M = 2L + 1,\\
\frac{L + 1/2 - w}{L}\sqrt{\l nP} & \text{ if }M = 2L.
\end{cases}
\label{X_1'}
\end{equation}
Because of the noisy feedback $\Yt_1$, the encoder can learn $Z_1 + \Zt_1 = \Yt_1 - X_1$. Subsequently it transmits
\[
X_i = (1 + \d)(Z_{i-1} + \Zt_{i-1}),\quad i \in[2: \eta],
\]
where  $\d > 0$ will be optimized later and the random time $\eta = \eta(w, Z^n, \Zt^n)$ is the largest $k \le \nb = \sqrt{n}$ such that
\[
 \sum_{i=1}^{k}X_i^2 \le nP.
\]

\smallskip\noindent{\bf Decoding.}
Upon receiving $Y^n$, the decoder estimates $X_1$ by
\[
\Xh_1 = \sum_{i=1}^{\nb}(-1)^{i-1}\frac{Y_i}{(1+\d)^{i-1}}
\]
and declares that $\wh$ is sent if
\[
\wh = \arg\min_{w\in [1: M]}|X_1(w) - \Xh_1|.
 \]

\begin{remark}

The main difference between this noisy feedback coding scheme and Pinsker's noise-free feedback coding scheme in the previous subsection is that we let the power of the initial transmission
grow linearly with the block length $n$ and thus that the initial transmission contains much more information about the message than in Pinsker's scheme.
This makes the coding scheme more robust to combat the noise in the feedback link.
\end{remark}

\smallskip\noindent{\bf Analysis of the probability of error.}
As before we assume that $\nb = \sqrt{n}$ is an integer.
Let
\begin{equation}
X_i' =
\begin{cases}
X_1  & \textrm{if $i = 1$},\\
(1 + \d)(Z_{i-1} + \Zt_{i-1})&\textrm{if $i \in [2: \nb]$},\\
0  &\textrm{otherwise},
\end{cases}
\end{equation}
and let $Y_i'$ and $\Xh_1'$ be defined as in $\eqref{Y_i'}$ and $\eqref{Xh_1'}$. Then, it can be easily shown that
\[
\Xh_1' = X_1 + (-1)^{\nb - 1}\frac{Z_{\nb}}{(1 + \d)^{\nb - 1}} + \sum_{i=1}^{\nb - 1}(-1)^i\frac{\Zt_i}{(1 + \d)^{i - 1}}.
\]
Thus we have
\begin{align*}
P_{\error}^{(n)}
&= \P\{W \ne \Wh\}\\
&\le \P\biggl\{|X_1 - \Xh_1| > \pam\biggr\} \\
&\le \P\biggl\{|X_1 - \Xh_1'| + |\Xh_1' - \Xh_1| > \pam\biggr\}\\
&\le \P\biggl\{|X_1 - \Xh_1'| > \pam\biggr\} + \P\{|\Xh_1' - \Xh_1| > 0\}\\
&=:P_1 + P_2.
\end{align*}
Now we upper bound the two terms. For the first term we have
\begin{align*}
P_1
&= \P\biggl\{\biggl\lvert(-1)^{\nb - 1}\frac{Z_{\nb}}{(1 + \d)^{\nb - 1}} + \sum_{i=1}^{\nb - 1}(-1)^i\frac{\Zt_i}{(1 + \d)^{i - 1}}\biggr\rvert > \pam\biggr\}\\
&= 2Q\bl\frac{\sqrt{\l nP/N}}{2L}\br \\
&\le \exp\bl-\frac{\l nP}{8L^2 N}\br,
\end{align*}
where
\begin{align*}
N
&=\sum_{i=1}^{\nb-1}\frac{\a}{(1+\d)^{2(i-1)}}+\frac{1}{(1+\d)^{2(\nb-1)}}\\
&= \frac{\a\bl 1-\frac{1}{(1+\d)^{2(\nb - 2)}}\br}{1-\frac{1}{(1+\d)^2}} + \frac{1}{(1+\d)^{2(\nb-1)}} \\
&\le \frac{\a(1+\d)^2}{(1+\d)^2 - 1} + \eps_n,
\end{align*}
where $\eps_n$ tends to zero as $n\to\infty$.
Thus
\begin{equation}
P_1 \le \exp\bl -\frac{\l nP}{8L^2}\bl  \frac{\a(1+\d)^2}{(1+\d)^2 - 1} + \eps_n\br^{-1}\br\label{P_1}.
\end{equation}
For the second term, we have
\begin{align*}
P_2
&\le \P\biggl\{\sum_{i=1}^{\nb}X_i^2 > nP\biggr\} \\
& \overset{(a)}{\le} \P\biggl\{\sum_{i=2}^{\nb}(1+\d)^2 (Z_{i-1} + \Zt_{i-1})^2> (1 - \l )nP\biggr\} \\
&= \P\biggl\{\chi_{\nb - 1}^2 > \frac{(1 - \l)nP}{(1 + \d)^2(1 + \a)}\biggr\},
\end{align*}
where $(a)$ follows since $X_1\le \l nP$ (recall $\eqref{X_1'}$). By $\eqref{Chi}$, we have
\begin{align}
P_2
&\le \P\biggl\{\chi_{\nb - 1}^2 > \frac{(1 - \l)nP}{(1 + \d)^2(1 + \a)}\biggr\}\nonumber\\
&\le \exp\bl -\frac{1}{2}\frac{(1 - \l)nP}{(1 + \d)^2(1 + \a)} + \frac{\nb - 1}{2}\log\frac{e(1 - \l)nP}{(\nb - 1)(1 + \d)^2(1 + \a)}\br\nonumber\\
&\le\exp\bl -\frac{1}{2}\frac{(1 - \l)nP}{(1 + \d)^2(1 + \a)} + n\eps_n\br\label{P_2},
\end{align}
where $\eps_n$ tends to zero as $n\to\infty$.
Therefore, the error exponent of the linear noisy feedback coding scheme is lower bounded as
\begin{align*}
E_M''(\a)
&= \limsup_{n\to\infty}- \frac{1}{n}\ln P_e^{(n)}\\
&= \limsup_{n\to\infty} -\frac{1}{n} \max\{\ln P_1, \ln P_2\}\\
&\ge \min \biggl\{\frac{\l P}{8L^2\a}\frac{(1 + \d)^2 - 1 }{(1 + \d)^2},\; \frac{(1 - \l)P}{2(1 + \d)^2(1 + \a)} \biggr\}.
\end{align*}
Now let
\[
\d = \d(\a) = \bl 1 + \sqrt{\frac{4L^2\a}{1+\a}}\br^{1/2} - 1
\]
and
\[
\l = \l(\a) = \bl 1 + \sqrt{\frac{1 + \a}{4L^2\a}}\br^{-1}.
\]
Then it can be readily verified that both terms in the minimum are the same and we have

\begin{equation}
E_M''(\a) \ge \frac{P}{2}\frac{1}{1 + \a + 4(\lfloor M/2 \rfloor)^2 \a + 4(\lfloor M/2 \rfloor)\sqrt{\a(1 + \a)}}\label{E_M''},
\end{equation}
which completes the proof of Theorem~$2$.
\begin{remark}
It is shown in Appendix~B that the lower bound in $\eqref{E_M''}$ is tight and characterizes the exact error exponent $E_M''(\a)$ of the linear noisy feedback coding scheme.

\end{remark}
\section{Discussion}
When $\a$ is very small, the linear feedback coding scheme (which is optimal for noise-free feedback) outperforms the two-stage (nonlinear) feedback coding scheme. When $\a$ is relatively large, however, linear feedback coding scheme amplifies the feedback noise, while the two-stage scheme achieves a more robust performance via signal protection. While this dichotomy agrees with the usual engineering intuition, it would be aesthetically more pleasing if a single feedback coding scheme performs uniformly better over all ranges of $\a$, and the search for such a coding scheme invites further investigation.
We finally note that $\a^* = 1/4$ is the threshold for all $M$ in the two-stage noisy feedback coding scheme (see Appendix~A). In both schemes, the error exponents are strictly larger than those for the no feedback case only when $\a$ is sufficiently small. Thus it is natural to ask
whether the noisy feedback is useful for all $\a$ or there exists a fundamental threshold beyond which noisy feedback becomes useless.

\bibliographystyle{IEEEtran}
\bibliography{nit} 

\appendices

\section{Proof of Theroem $1$ for the General Case}
\smallskip\noindent{\bf Encoding.}
In stage $1$, the encoder uses the simplex signaling for an $M$-ary message:
\[
x^{\l n}(w) = A\bl e_w - \frac{1}{M}\sum_{w = 1}^M e_w\br\quad\text{for }w\in [1: M],
\]
where $A = \sqrt{M\l nP/(M - 1)}$ and
\[
e_w = (\underbrace{0, \cdots, 0}_{w - 1}, 1, 0, \cdots, 0).
\]
Then based on the noisy feedback $\yt^{\l n}$, the encoder chooses the two most probable message estimates $\wt_1$ and $\wt_2$ among $M$ candidates. In stage $2$, the encoder uses antipodal signaling for $w$ if $w\in \{\wt_1, \wt_2\}$ and transmits all-zero sequence otherwise.

\smallskip\noindent{\bf Decoding.}
The signal protection region for the $M$-ary message is defined as in  \eqref{protection} (with $w$, $w'$, $w''$$\in$ $[1: M]$).
The decoder makes a decision immediately at the end of stage $1$ if the received signal $y^{\l n}$ lies in one of the signal protection regions.
Otherwise, it chooses the two most probable message estimates $\wh_1$ and $\wh_2$, and wait for the transmission in stage $2$.
At the end of stage $2$, the decoder declares that $\wh$ is sent if
\[
\wh = \arg\min_{w \in \{\wh_1, \wh_2\}} \bigl( ||x^{\l n}(w) - y^{\l n}||^2 + ||x^{n}_{\l n + 1}(w) - y_{\l n + 1}^n||^2\bigr)^{1/2}.
\]

\smallskip\noindent{\bf Analysis of the probability of error.}
Let ($\Wt_1$, $\Wt_2$) and ($\Wh_1$, $\Wh_2$) denote the pairs of the two most probable message estimates at the encoder and the decoder, respectively.
The decoder makes an error only if one or more of the following events occur:

\begin{itemize}
\item decoding error at the end of stage $1$
\[
\Ec_1 = \blb Y^{\l n}\in \cup_{w \ne 1} B_w\cup (\cup_{w, w'\ne 1} A_{ww'}')\brb,
\]
\item miscoordination due to the feedback noise
\begin{align*}
\tilde{\Ec}_{1w} &= \blb Y^{\l n}\in A_{1w}'\text{ and } \Yt^{\l n}\in \cup_{\{w', w''\}\ne\{1, w\}} A_{w'w''}\brb,
\end{align*}
\item decoding error at the end of stage $2$
\[
\Ec_2 = \blb W\in\{\Wh_1, \Wh_2\} = \{\Wt_1, \Wt_2\} \text{ and }\Wh \ne W\brb.
\]
\end{itemize}
Thus, the probability of error is upper bounded as
\[
P_e^{(n)} \le \P(\Ec_1) + M\P(\tilde{\Ec}_{1w}) + \P(\Ec_2).
\]

As before, we assume that $W = 1$ was sent.
For the first term, by the union of events bound,
\begin{align*}
\P(\Ec_1)
&= \P\blb Y^{\l n}\in \cup_{w \ne 1} B_w\cup (\cup_{w, w'\ne 1} A_{ww'}')\brb\\
&\le M^2\P\blb Y^{\l n}\in B_2\cup A_{23}'\brb.
\end{align*}
For $\P(\tilde{\Ec}_{1w})$, again by the union of events bound,
\begin{align*}
\P(\tilde{\Ec}_{1w})
&= \P\blb Y^{\l n}\in A_{1w}'\text{ and } \Yt^{\l n}\in \cup_{\{w', w''\}\ne\{1, w\}} A_{w'w''}\brb\\
&\le M^2\P\blb Y^{\l n}\in A_{1w}'\text{ and } \Yt^{\l n}\in A_{w'w''}\brb.
\end{align*}
We use $d_j'$, $j \in [1: 6]$, to denote the distances corresponding
to $d_j$ in the $M = 3$ case (see Fig.~\ref{fig:error_events}). It can be easily checked that $d_j' = d_j\sqrt{3(M - 1)/(2M)}$.
Thus by replacing $d_5$ by $d_5'$ in \eqref{d_5} and $d_6$ by $d_6'$ in \eqref{d_6}, we have
\[
\P(\Ec_1)\le M^2Q(d_5') \le \frac{M^2}{2}\exp\biggl(-\frac{M}{12(M - 1)}\l nP(s^2 - 2s + 4)\biggr)
\]
and
\[
\P(\tilde{\Ec}_{12})\le M^2Q\bl \frac{d_6'}{\sqrt{\a}}\br \le \frac{M^2}{2}\exp\biggl(-\frac{s^2M}{16(M - 1)\a}\l nP\biggr).
\]
The third term $\P(\Ec_2)$ can be upper bounded in the same manner as for the $M = 3$ case,
\begin{align*}
\P(\Ec_2)
&= Q \bl-\sqrt{\bl1-\frac{M-2}{2(M-1)}\lambda\br nP}\br \\
&\le \frac{1}{2}\exp\bl\frac{nP}{2}\bl 1 - \frac{M - 2}{2(M - 1)}\l\br\br.
\end{align*}
Therefore,
\begin{align*}
E_M'(\a)
&= \limsup_{n\to\infty}- \frac{1}{n}\ln P_e^{(n)}\\
&\ge \limsup_{n\to\infty} -\frac{1}{n} \max\{\ln \P(\Ec_1), \ln \P(\tilde{\Ec}_{12}), \ln \P(\Ec_2)\}\\
&\ge \min \biggl\{\frac{\l MP}{12(M-1)}(s^2 - 2s + 4),\;
\frac{s^2\l MP}{16(M-1)\a},\;
\frac{P}{2}\bl 1 - \frac{M - 2}{2(M - 1)}\l\br
\biggr\}.
\end{align*}
Now let
\[
\a = \a^*(s) = \frac{3s^2}{4(s^2 - 2s + 4)}
\]
and
\[
\l = \l^*(s) = \bl\frac{M}{6(M - 1)}(s^2 - 2s + 4) + \frac{M - 2}{2(M - 1)}\br^{-1}.
\]
Then it can be readily verified that all the three terms in the minimum are the same and we have
\[
E_M'(\a^*(s)) \ge \frac{P}{2}\bl 1 - \frac{3(M - 2)}{M(s^2 - 2s + 4) + 3(M - 2)}\br=: \phi(s).
\]
Note that if $s < 1$,
\[
\phi(s)> \frac{M}{4(M - 1)}P = E_M(\infty),
\]
and $\a^*(s)$ is
monotonically increasing over $s\in [0, 1]$. Thus
\[
E_M'(\a) > E_M(\infty)\quad\text{for } \a < \a^*(1) = \frac{1}{4}.
\]
This completes the proof of Theorem $1$ for the general case.

\begin{remark}
Note that $E_M'(\a)$ is decreasing in $M$, while $\a^*(s)$ is still independent of $M$.
\end{remark}
\section{$E_M''(\a)$ is Tight}
Now we show that $E_M''(\a)$ is the exact error exponent for the linear noisy feedback coding scheme. Consider
\begin{align}
P_{\error}^{(n)}
&=\P\{W \ne \Wh\}\nonumber\\
&\ge A\P\{W \ne \Wh | W \in [2: M-1]\}\nonumber\\
&= A\P\biggl\{|X_1 - \Xh_1| > \pam\biggr\}\nonumber\\
&= A\P\biggl\{ \bigl|(X_1 - \Xh_1') - (\Xh_1' -\Xh_1)\bigr| > \pam\biggr\}\nonumber\\
&\ge A\P\biggl\{ |X_1 - \Xh_1'| > \pam \text{ and } |\Xh_1' -\Xh_1| = 0\biggr\}\nonumber\\
&\overset{(a)}{\ge} A\P\biggl\{ |X_1 - \Xh_1'| > \pam \text{ and } \sum_{i=1}^{\nb}X_i^2 \le nP\biggr\}\nonumber\\
&\ge A\bl\P\biggl\{ |X_1 - \Xh_1'| > \pam\biggr\} + \P \biggl\{\sum_{i=1}^{\nb}X_i^2 \le nP\biggr\} - 1\br\nonumber\\
&= A\bl\P\biggl\{ |X_1 - \Xh_1'| > \pam\biggr\} + \P \biggl\{\sum_{i=1}^{\nb}X_i^2 > nP\biggr\}\br\label{tight},
\end{align}
where $A = (M - 2)/M$ and $(a)$ follows since $\sum_{i=1}^{\nb}X_i^2 \le nP$ implies $|\Xh_1' - \Xh| = 0$.
Combining $\eqref{tight}$ with the fact that the upper bounds $\eqref{P_1}$ and $\eqref{P_2}$ are both tight in exponent, we conclude that $E_M''(\a)$ is the exact error exponent for the linear noisy feedback coding scheme.

\end{document}